\documentstyle{mn}

\newif\ifAMStwofonts

\newif\ifAMStwofonts

\ifoldfss
  \ifCUPmtlplainloaded \else
    \NewTextAlphabet{textbfit} {cmbxti10} {}
    \NewTextAlphabet{textbfss} {cmssbx10} {}
    \NewMathAlphabet{mathbfit} {cmbxti10} {} 
    \NewMathAlphabet{mathbfss} {cmssbx10} {} 
  \fi
  \ifAMStwofonts
    \ifCUPmtlplainloaded \else
      \NewSymbolFont{upmath} {eurm10}
      \NewSymbolFont{AMSa} {msam10}
      \NewMathSymbol{\upi}     {0}{upmath}{19}
      \NewMathSymbol{\umu}     {0}{upmath}{16}
      \NewMathSymbol{\upartial}{0}{upmath}{40}
      \NewMathSymbol{\leqslant}{3}{AMSa}{36}
      \NewMathSymbol{\geqslant}{3}{AMSa}{3E}

      \let\leq=\leqslant 
       
    \fi
  \fi
\fi 

\ifnfssone
  \newmathalphabet{\mathit}
  \addtoversion{normal}{\mathit}{cmr}{m}{it}
  \addtoversion{bold}{\mathit}{cmr}{bx}{it}
  \newmathalphabet{\mathbfit} 
  \addtoversion{normal}{\mathbfit}{cmr}{bx}{it}
  \addtoversion{bold}{\mathbfit}{cmr}{bx}{it}
  \newmathalphabet{\mathbfss} 
  \addtoversion{normal}{\mathbfss}{cmss}{bx}{n}
  \addtoversion{bold}{\mathbfss}{cmss}{bx}{n}
  \ifAMStwofonts
    \ifCUPmtlplainloaded \else
      \UseAMStwoboldmath
      \makeatletter
      \new@mathgroup\upmath@group
      \define@mathgroup\mv@normal\upmath@group{eur}{m}{n}
      \define@mathgroup\mv@bold\upmath@group{eur}{b}{n}
      \edef\UPM{\hexnumber\upmath@group}
      \new@mathgroup\amsa@group
      \define@mathgroup\mv@normal\amsa@group{msa}{m}{n}
      \define@mathgroup\mv@bold\amsa@group{msa}{m}{n}
      \edef\AMSa{\hexnumber\amsa@group}
      \makeatother
      \mathchardef\upi="0\UPM19
      \mathchardef\umu="0\UPM16
      \mathchardef\upartial="0\UPM40
      \mathchardef\leqslant="3\AMSa36
      \mathchardef\geqslant="3\AMSa3E

      \let\leq=\leqslant 

    \fi
  \fi
\fi 

\ifnfsstwo
  \DeclareMathAlphabet{\mathbfit}{OT1}{cmr}{bx}{it}
  \SetMathAlphabet\mathbfit{bold}{OT1}{cmr}{bx}{it}
  \DeclareMathAlphabet{\mathbfss}{OT1}{cmss}{bx}{n}
  \SetMathAlphabet\mathbfss{bold}{OT1}{cmss}{bx}{n}
  \ifAMStwofonts
    \ifCUPmtlplainloaded \else
      \DeclareSymbolFont{UPM}{U}{eur}{m}{n}
      \SetSymbolFont{UPM}{bold}{U}{eur}{b}{n}
      \DeclareSymbolFont{AMSa}{U}{msa}{m}{n}
      \DeclareMathSymbol{\upi}{0}{UPM}{"19}
      \DeclareMathSymbol{\umu}{0}{UPM}{"16}
      \DeclareMathSymbol{\upartial}{0}{UPM}{"40}
      \DeclareMathSymbol{\leqslant}{3}{AMSa}{"36}
      \DeclareMathSymbol{\geqslant}{3}{AMSa}{"3E}

      \let\leq=\leqslant 

    \fi
  \fi
\fi 

\ifCUPmtlplainloaded \else
  \ifAMStwofonts \else 
    \def\upi{\pi}
    \def\umu{\mu}
    \def\upartial{\partial}
  \fi
\fi

\title[Observations of 4U 1630-47]
{RXTE observations of 4U~1630--47 during the peak of its 1998 outburst}
\author[S.Trudolyubov, K.Borozdin and W.Priedhorsky]
{Sergey P.~Trudolyubov $^{1,2}$\thanks{E-mail:tsp@hea.iki.rssi.ru}, 
  Konstantin N.~Borozdin $^{2,1}$ and
  William C.~Priedhorsky$^{2}$
\\
  $^1$ Space Research Institute, RAS, Profsoyuznaya 84/32, 117810 Moscow, 
	Russia \\
  $^2$ MS D436, Los Alamos National Laboratory, Los Alamos,
              87545 New Mexico, USA
      }

\date{Accepted       Received       in original form ??}

\pagerange{\pageref{firstpage}--\pageref{lastpage}}
\pubyear{1999}

\input epsf

\begin{document}

\maketitle

\label{firstpage}

\begin{abstract}
\large
We present an analysis of the RXTE observations of 
4U~1630--47 during its outburst of 1998. The light curve 
and the spectral evolution of the outburst were distinctly 
different from the outbursts of the same source in 1996 
and in 1999. Special emphasis of our analysis was on the 
observations taken during the initial rise of the flux and 
during the maximum of the outburst.  The maximum of the outburst 
was divided into three plateaus, with almost constant flux 
within each plateau, and fast jumps between them.
The spectral and timing parameters are stable for each 
individual plateau, but distinctly different between the plateaus.  
The variability detected on the first plateau is of special interest.
During these observations the source exhibits quasi-regular modulations 
with period of $\sim 10 - 20$ s.
Our analysis revealed significant differences in spectral and temporal 
behavior of the source at high and low fluxes during this period of time.
The source behavior can be generally explained 
in the framework of the two--phase model of the accretion flow, involving 
a hot inner comptonization region and surrounding optically thick disk.

The variability and spectral evolution of the source were 
similar to what was observed earlier for other X-ray Novae. 
We show that 4U~1630--47 has a variety of properties which 
are typical for Galactic black hole binaries, both transient 
and persistent. We argue that this system may be an intermediate 
case between different groups of black hole candidates.

\end{abstract}


\large
\section{INTRODUCTION}

Recurrent outbursts of X-ray transient source 4U~1630--47 
have been observed by X-ray experiments for thirty years.
The first known outburst was recorded by Vela-5B in 1969 
\cite{pried86}. Multiple outbursts were detected with various 
all-sky monitors in later years.  The outbursts occurred quasi-
regularly, with a recurrence period of $\sim$ 600 days 
\cite{jones76,pried86,parmar95}, however, a strict 
regularity was ruled out by observations of several out-of-phase 
outbursts \cite{kal78,mccol99}.

\begin{figure}
\epsfxsize=8.5cm
\epsffile{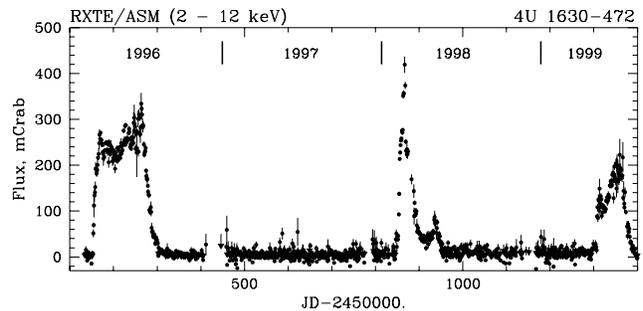}
\caption{The long-term X-ray flux history of 4U~1630--47 according to 
the data of RXTE/ASM observations (each point represents daily 
averaged value of the X-ray flux). \label{lc_general}}
\end{figure}

4U~1630--47 belongs to the group of X-ray transients known as X-ray 
Novae \cite{sun94,tsh96}. All sources in this class are assumed 
to be recurrent, but only a few have been observed in more than one 
X-ray outburst. The typical recurrence time for outbursts of such sources 
is $\sim$10--50 years, and 4U~1630--47, with its relatively frequent 
outbursts is therefore unusual. Its X-ray spectral and timing behavior
during outbursts is however quite typical for black hole binaries with
low mass companions \cite{tl95}.  No optical counterpart is known
for 4U~1630--47, probably due to its high reddening and crowded field
\cite{parmar86}.  Its identification as a black hole candidate in
a low mass binary, derived from X-ray observations, looks nevertheless 
quite reliable.

The short recurrence time of the outbursts makes the source a convenient
choice for studies of black hole X-ray transient behavior. In this 
paper we present an analysis of the observations of 4U~1630--47 during 
its 1998 outburst based on RXTE data. Our special emphasis was on the 
maximum of the outburst.

\section{OBSERVATIONS AND DATA ANALYSIS}

\subsection{RXTE observations}

RXTE satellite \cite{bsr93} performed the extended series of 
pointed observations during 1998 outburst of 4U~1630--47. The observations 
started during the rise of the flux and followed the evolution of the source 
deep into the decline of the outburst. Total number of the observations was 
$\sim$ 100. In the Table \ref{obslog} we listed observations we analyzed 
in detail. 

\begin{table}
\small
\caption{The list of RXTE/PCA observations of 
4U 1630--47 used in the analysis. 
\label{obslog}} 
\begin{tabular}{cccccc}
\hline
$\#$ & Obs.ID & TJD$^{a}$ & Date & Start & Exp.$^{b}$\\
     &        &           & (UT) & (UT)  & (s)       \\
\hline
1 & 30178-01-01-00 & 853.06 & 09/02/98 & 01:22 & 1034 \\
2 & 30178-01-02-00 & 853.66 & 09/02/98 & 15:44 & 1134 \\
3 & 30178-02-01-00 & 855.05 & 11/02/98 & 01:07 & 3554 \\
4 & 30188-02-01-00 & 855.43 & 11/02/98 & 10:17 & 1175 \\
5 & 30178-01-03-00 & 855.84 & 11/02/98 & 20:11 & 1438 \\
6 & 30188-02-02-00 & 856.12 & 12/02/98 & 02:46 & 1846 \\
7 & 30178-01-04-00 & 856.63 & 12/02/98 & 15:03 & 1659 \\
8 & 30188-02-03-00 & 856.65 & 12/02/98 & 15:38 & 1509 \\
9 & 30178-02-01-01 & 856.87 & 12/02/98 & 20:51 & 2919 \\
10& 30188-02-04-00 & 856.98 & 12/02/98 & 23:26 & 2015 \\
11& 30178-01-05-00 & 857.12 & 13/02/98 & 02:47 & 1441 \\
12& 30188-02-05-00 & 857.71 & 13/02/98 & 17:01 & 2266 \\
13& 30178-02-02-00 & 857.79 & 13/02/98 & 19:02 & 6540 \\
14& 30188-02-06-00 & 858.05 & 14/02/98 & 01:05 & 2168 \\
15& 30178-01-06-00 & 858.70 & 14/02/98 & 16:40 & 1453 \\
16& 30188-02-07-00 & 858.72 & 14/02/98 & 17:10 & 1786 \\
17& 30178-02-02-01 & 858.77 & 14/02/98 & 18:31 & 8414 \\
18& 30178-01-07-00 & 859.84 & 15/02/98 & 20:04 & 1370 \\
19& 30188-02-08-00 & 860.12 & 16/02/98 & 02:48 & 1828 \\
20& 30188-02-09-00 & 860.56 & 16/02/98 & 13:29 & 1597 \\
21& 30178-01-08-00 & 860.72 & 16/02/98 & 17:20 & 1180 \\
22& 30178-01-09-00 & 861.69 & 17/02/98 & 16:38 & 1314 \\
23& 30188-02-10-00 & 861.72 & 17/02/98 & 17:21 & 1620 \\
24& 30178-01-10-00 & 862.65 & 18/02/98 & 15:41 & 1638 \\
25& 30188-02-11-00 & 862.71 & 18/02/98 & 16:54 & 2864 \\
26& 30178-02-03-00 & 862.77 & 18/02/98 & 18:22 & 9296 \\
27& 30178-01-11-00 & 863.70 & 19/02/98 & 16:41 & 1237 \\ 
28& 30188-02-12-00 & 863.71 & 19/02/98 & 17:06 & 2245 \\
29& 30188-02-13-00 & 864.19 & 20/02/98 & 04:26 & 1233 \\
30& 30188-02-14-00 & 864.32 & 20/02/98 & 07:41 & 1080 \\
31& 30178-01-12-00 & 864.63 & 20/02/98 & 15:07 & 1579 \\
32& 30188-02-15-00 & 865.05 & 21/02/98 & 01:06 & 1510 \\
33& 30188-02-16-00 & 865.32 & 21/02/98 & 07:41 & 912  \\
34& 30188-02-17-00 & 866.64 & 22/02/98 & 15:21 & 1419 \\
35& 30178-01-13-00 & 867.50 & 23/02/98 & 11:56 & 1524 \\
36& 30188-02-18-00 & 868.57 & 24/02/98 & 13:38 & 3620 \\
37& 30188-02-19-00 & 869.65 & 25/02/98 & 15:37 & 1698 \\
38& 30188-02-20-00 & 870.18 & 26/02/98 & 04:21 & 900  \\
39& 30178-01-14-00 & 870.98 & 26/02/98 & 23:28 & 1895 \\
40& 30188-02-21-00 & 871.26 & 27/02/98 & 06:11 & 2175 \\
41& 30188-02-21-01 & 871.41 & 27/02/98 & 09:50 & 772  \\ 
42& 30188-02-22-00 & 872.34 & 28/02/98 & 08:12 & 4972 \\ 
43& 30188-02-23-00 & 873.14 & 01/03/98 & 03:19 & 1347 \\ 
44& 30178-01-15-00 & 874.52 & 02/03/98 & 12:23 & 1400 \\
45& 30178-01-16-00 & 881.58 & 09/03/98 & 13:55 & 1875 \\
46& 30178-01-17-00 & 884.94 & 12/03/98 & 22:33 & 1590 \\       
47& 30178-01-18-00 & 889.72 & 17/03/98 & 17:08 & 1359 \\
48& 30172-01-05-00 & 891.65 & 19/03/98 & 15:30 & 9152 \\
\hline
\end{tabular}
\begin{list}{}{}
\item[$^a$]-- Truncated Julian Date: TJD=JD-2450000.
\item[$^b$]-- Dead time corrected value of the PCA exposure
\end{list}
\end{table}

For the processing of the PCA and HEXTE data we used standard RXTE FTOOLS 
version 4.2 tasks.

\begin{figure}
\epsfxsize=8.5cm
\epsffile{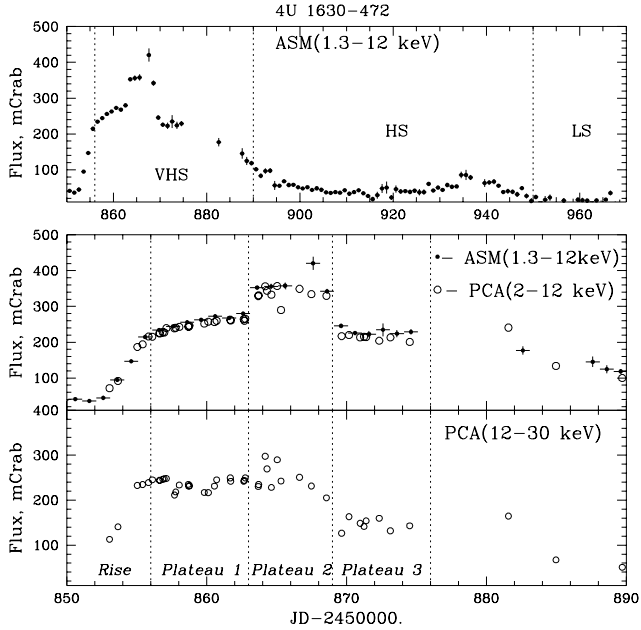}
\caption{The light curves of 4U~1630--47 during its 1998 outburst 
according to the data of RXTE/PCA and ASM instruments. 
In upper panel general evolution of the source flux during 
the whole outburst is shown ($1.3 - 12$ keV energy band, ASM data). 
The detailed flux histories during the initial rise and maximum flux 
are shown in middle ($1.3 - 12$ keV energy band, ASM data -- 
filled circles; $2 -12$ keV energy range, PCA data -- 
hollow circles) and lower ($12 - 30$ keV energy 
range, PCA data -- hollow circles) panels.  \label{lightcurve}}
\end{figure}

\subsection{Spectral analysis}

\begin{figure*}
\hbox{
\epsfxsize=9.0cm
\epsffile{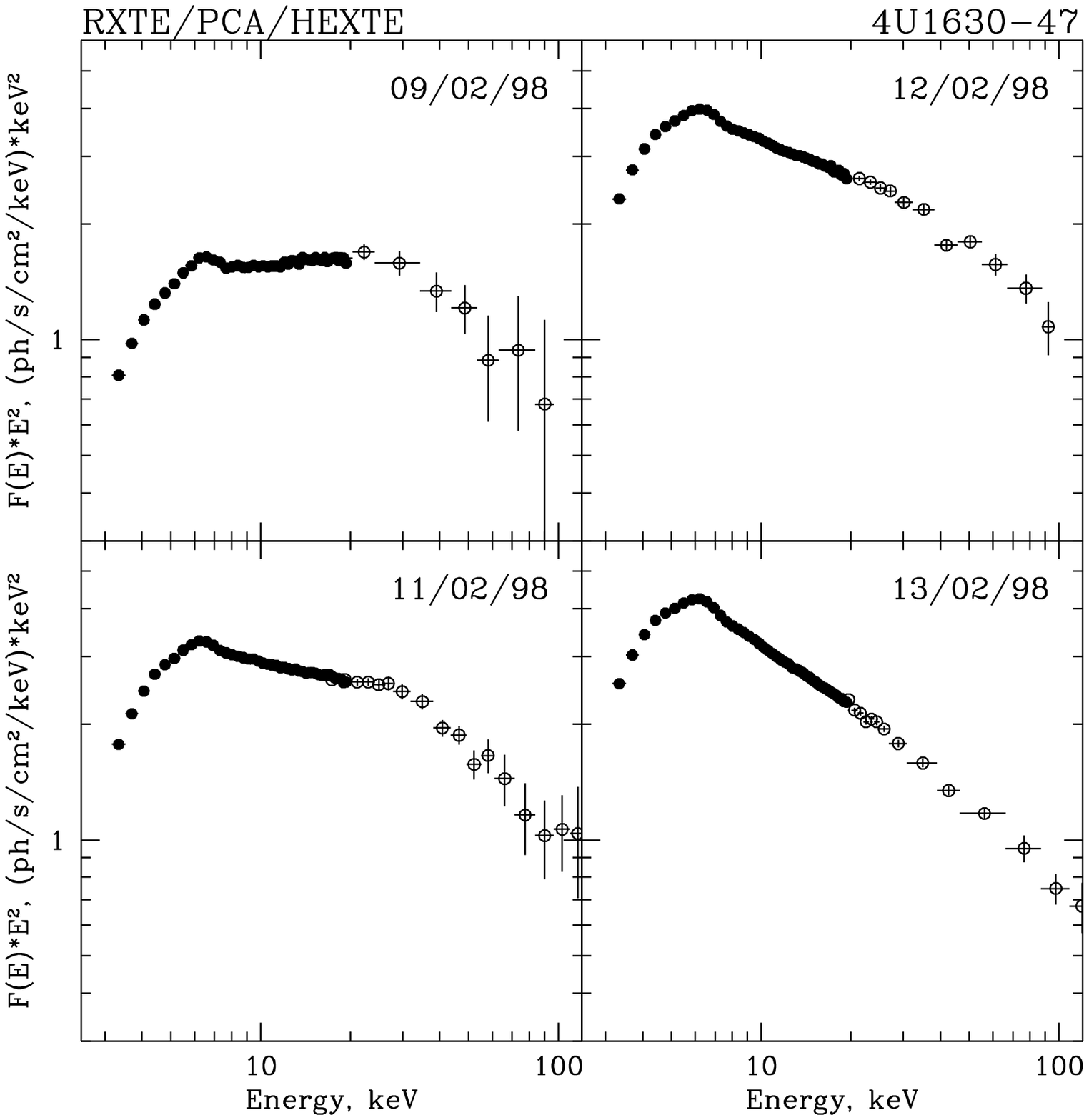}
\epsfxsize=9.0cm
\epsffile{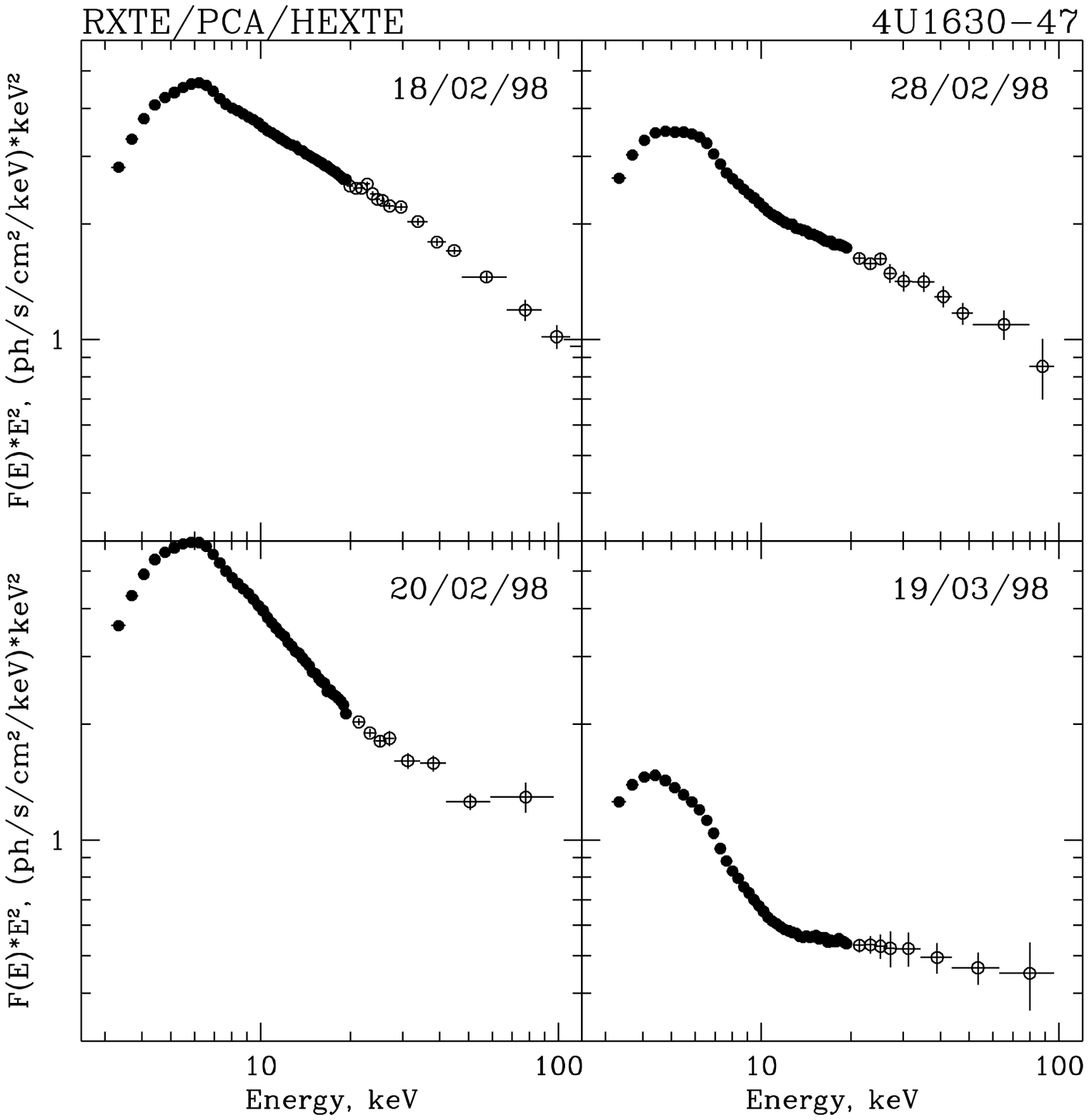}
}
\caption{Typical broad-band spectra of 4U1630-47 during the initial 
rise and maximum phase of the 1998 outburst. Filled and 
hollow circles represent the data of PCA and HEXTE instruments
respectively. \label{spectra}}
\end{figure*}

For the spectral analysis we used PCA data collected in the $3 - 20$ keV 
energy range. PCA response matrixes for individual observations were 
constructed using PCARMF v3.5, background estimation was performed applying 
a Very-Large Events (VLE)--based model. To evaluate correct source flux, the 
standard dead time correction procedure was applied to the PCA data. In 
order to account for the uncertainties of the response matrix, a 1$\%$ -- 
systematic error was added to the statistical error for each PCA energy 
channel.
   
We used the HEXTE response matrices, released on April 3, 1997 and standard 
off--source observations for each cluster of detectors to subtract background. 
In order to account for the uncertainties in the response and background 
determination, only data in the $19 - 100$ keV energy range were taken for 
the spectral analysis.

We generated the energy spectra of 4U~1630--47 averaging the data over 
the whole single observations using the {\em Standard2} mode data. To 
trace the evolution of the source spectral properties on the short time 
scales the data in the {\em Binned} and {\em Event} modes were used.  

As we are interested in the general character of the 4U~1630-47 
spectral evolution, we used only simplest models to approximate its 
spectra. Due to the cross--calibration uncertainties the data of PCA and 
HEXTE instruments were analyzed separately. For the approximation of the 
PCA spectral data a sum of the XSPEC ``multicolor disk black body'' model 
\footnote{Note 
that no corrections for the electron scattering and effects of general 
relativity were made \cite{ss73,sht95}. In addition, this model assumes 
incorrect radial dependence of the disk effective temperature within 
$\sim 5$ gravitational radii. Thus the inferred value of the effective 
radius ${\rm R_{in}}$ should not be treated as the actual value of the 
characteristic radius of the optically thick emitting region} and a power 
law model, corrected for the interstellar absorption were used. Although 
this model provides 
satisfactory description to the overall shape of the spectrum, sometimes 
there is a significant excess of emission in the $6 - 8$ keV region 
(with an average ratio of excess to the model continuum of $\sim 1 - 3 \%$), 
which is attributed to the presence of the iron emission/absorption 
complex. Due to the relatively low energy resolution of the PCA instrument 
(${\rm \Delta E \sim 1}$ keV in the $6 - 8$ keV range) it is not possible to 
carry out a detailed analysis of this spectral feature. The presence of 
the iron emission/absorption features shows the evidence for the 
additional 'reprocessed' component in the source spectrum, 
but its approximation by certain type of model requires further physical 
justification, so we have decided not to include 'reprocessed' component 
to the fit. \footnote{The inclusion of Gaussian emission line and 
absorption edge model features has minor effect on the values of the 
main model parameters and integrated energy flux, derived through the spectral 
fitting, except for the value of color temperature T${\rm _{d}}$ of the soft 
component, which decreases by $\sim 20 \%$}. For the analytic approximation 
of the HEXTE data a simple power model was used.

\begin{figure}
\epsfxsize=8.5cm
\epsffile{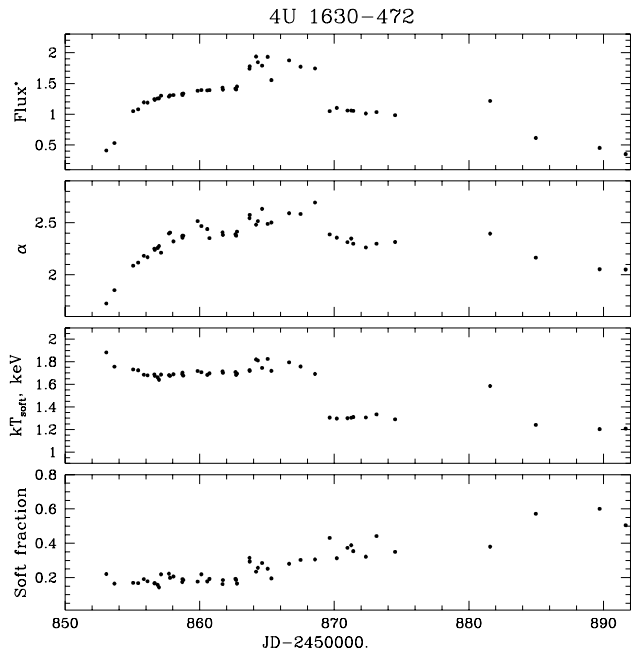}
\caption{The evolution of the spectral parameters of the multicolor
disk black body plus power law model approximation for energy spectra 
of 4U1630-47 during the 1998 outburst. 'Soft fraction' denotes the the
ratio of the flux of component to the total flux from the source in
the $3 - 20$ keV energy range. 'Flux'$^{*}$ denotes the total model flux 
in the $3 - 20$ keV energy range corrected for the low energy absorption 
(in units of $\times 10^{-8}$ erg/s/cm$^{2}$).  \label{spec_evol}}
\end{figure}

Typical examples of the broad--band energy spectra of 4U~1630--47 in units 
of ${\rm E^{2} \times F(E)}$ (keV$^{2}$ phot/keV) are shown in Fig. 
\ref{spectra}. The results of fitting of the PCA and HEXTE spectral data 
with analytical 
models described above are presented in Table \ref{params}. Figure 
\ref{spec_evol} shows the evolution of fitting parameters with time.

\begin{table*}
\small
\caption{Spectral parameters of 4U~1630-47, derived using the
combination of the multicolor disc blackbody (Mitsuda et al. 1984) 
and power law models with correction to the interstellar absorption 
(PCA  data, $3 - 20$ keV energy range). Parameter errors correspond 
to $1 \sigma$ confidence level for the assumed $1 \%$ systematic 
uncertainty of data. For the approximation of HEXTE data in the 
$19 - 100$ keV energy range a simple power law model was used. 
In order to improve statistical significance the data of HEXTE 
instrument were averaged over several observations corresponding 
to the same date. 
\label{params}}  
\begin{tabular}{rcccccccccc}
\hline
$\#$ & ${\rm T_{d}}$, keV & ${\rm Norm^{a}}$ & ${\rm f_{soft}^{bol}}$$^{b}$ & $\alpha_{\rm pl}$ & 
${\rm N_{\rm H}^{c}}$ & ${\rm f_{total}^{d}}$ & ${\rm f_{\rm soft}^{e}}$ & $\chi^2$(d.o.f.) & $\alpha_{\rm pl}^{\rm f}$\\
\hline
1 &$1.88\pm0.06$&$6.0\pm0.7$ &$1.63$&$1.73\pm0.06$&$7.22\pm0.37$ &$4.13$ &$0.91$&63.5(40) & $2.46\pm0.07$\\
2 &$1.76\pm0.06$&$8.1\pm1.2$ &$1.66$&$1.85\pm0.05$&$7.48\pm0.35$ &$5.32$ &$0.88$&59.5(40) & $2.55\pm0.13$\\
3 &$1.73\pm0.05$&$17.6\pm2.0$&$3.42$&$2.09\pm0.04$&$7.84\pm0.31$ &$10.50$&$1.78$&21.7(40) & $2.51\pm0.04$\\
4 &$1.72\pm0.06$&$18.2\pm2.3$&$3.48$&$2.12\pm0.05$&$7.71\pm0.35$ &$10.80$&$1.81$&30.4(40) & $$\\
5 &$1.68\pm0.04$&$25.9\pm2.9$&$4.49$&$2.18\pm0.04$&$8.05\pm0.32$ &$11.94$&$2.29$&15.7(40) & $$\\
6 &$1.68\pm0.05$&$24.3\pm2.7$&$4.17$&$2.17\pm0.04$&$7.78\pm0.32$ &$11.88$&$2.12$&17.7(40) & $2.51\pm0.03$\\
7 &$1.69\pm0.05$&$23.3\pm2.8$&$4.08$&$2.25\pm0.04$&$8.08\pm0.33$ &$12.47$&$2.08$&26.5(40) & $$\\
8 &$1.68\pm0.04$&$24.0\pm3.0$&$4.09$&$2.24\pm0.04$&$7.90\pm0.33$ &$12.33$&$2.07$&23.7(40) & $$\\
9 &$1.66\pm0.05$&$23.9\pm2.1$&$3.94$&$2.26\pm0.03$&$7.97\pm0.31$ &$12.57$&$1.98$&25.4(40) & $$\\
10&$1.64\pm0.05$&$23.2\pm3.3$&$3.63$&$2.28\pm0.03$&$8.13\pm0.32$ &$12.60$&$1.80$&19.0(40) & $$\\
11&$1.69\pm0.03$&$32.0\pm3.0$&$5.60$&$2.21\pm0.04$&$7.88\pm0.32$ &$13.01$&$2.85$&21.2(40) & $2.71\pm0.02$\\
12&$1.68\pm0.03$&$32.8\pm3.0$&$5.65$&$2.40\pm0.04$&$8.35\pm0.33$ &$12.88$&$2.87$&20.2(40) & $$\\
13&$1.68\pm0.03$&$30.1\pm2.9$&$5.13$&$2.41\pm0.03$&$8.38\pm0.32$ &$13.06$&$2.60$&18.6(40) & $$\\
14&$1.69\pm0.04$&$30.1\pm2.9$&$5.30$&$2.32\pm0.04$&$8.03\pm0.33$ &$13.12$&$2.71$&21.3(40) & $2.70\pm0.02$\\
15&$1.69\pm0.04$&$25.8\pm3.2$&$4.53$&$2.37\pm0.04$&$8.22\pm0.34$ &$13.30$&$2.31$&22.7(40) & $$\\
16&$1.70\pm0.04$&$26.8\pm3.0$&$4.87$&$2.36\pm0.04$&$8.03\pm0.34$ &$13.12$&$2.50$&23.9(40) & $$\\
17&$1.68\pm0.03$&$28.7\pm2.9$&$4.90$&$2.38\pm0.03$&$8.23\pm0.30$ &$13.37$&$2.48$&17.1(40) & $$\\
18&$1.72\pm0.03$&$25.1\pm3.3$&$4.72$&$2.52\pm0.04$&$8.83\pm0.36$ &$13.81$&$2.44$&26.4(40) & $2.57\pm0.06$\\
19&$1.71\pm0.03$&$32.2\pm3.3$&$5.92$&$2.47\pm0.04$&$8.56\pm0.35$ &$13.92$&$3.05$&19.8(40) & $2.70\pm0.06$\\
20&$1.68\pm0.04$&$28.0\pm3.3$&$4.85$&$2.44\pm0.04$&$8.31\pm0.34$ &$13.87$&$2.47$&26.4(40) & $$\\
21&$1.70\pm0.04$&$29.2\pm3.2$&$5.22$&$2.35\pm0.04$&$7.81\pm0.35$ &$13.91$&$2.68$&26.8(40) & $$\\
22&$1.72\pm0.04$&$24.0\pm3.3$&$4.48$&$2.41\pm0.04$&$8.32\pm0.36$ &$14.34$&$2.32$&21.6(40) & $2.57\pm0.07$\\
23&$1.70\pm0.04$&$28.0\pm3.1$&$5.07$&$2.38\pm0.04$&$8.02\pm0.34$ &$14.00$&$2.60$&19.7(40) & $$\\
24&$1.71\pm0.04$&$28.9\pm3.1$&$5.31$&$2.39\pm0.04$&$8.02\pm0.34$ &$14.17$&$2.73$&22.9(40) & $2.60\pm0.02$\\
25&$1.68\pm0.04$&$29.6\pm3.1$&$5.14$&$2.38\pm0.04$&$8.10\pm0.33$ &$14.03$&$2.62$&21.2(40) & $$\\
26&$1.69\pm0.04$&$26.3\pm3.0$&$4.69$&$2.42\pm0.04$&$8.38\pm0.32$ &$14.51$&$2.40$&19.8(40) & $$\\
27&$1.73\pm0.02$&$55.0\pm3.9$&$10.53$&$2.54\pm0.05$&$8.70\pm0.36$&$17.39$&$5.48$&19.7(40) & $2.76\pm0.04$\\
28&$1.72\pm0.02$&$53.2\pm3.8$&$10.05$&$2.57\pm0.04$&$8.97\pm0.35$&$17.77$&$5.21$&21.0(40) & $$\\
29&$1.82\pm0.03$&$35.2\pm3.7$&$8.33$&$2.48\pm0.05$&$8.77\pm0.37$ &$19.37$&$4.53$&27.1(40) & $2.50\pm0.07$\\
30&$1.81\pm0.03$&$37.6\pm3.7$&$8.76$&$2.52\pm0.05$&$8.69\pm0.37$ &$18.46$&$4.75$&21.0(40) & $$\\
31&$1.75\pm0.02$&$48.4\pm4.0$&$9.70$&$2.63\pm0.05$&$9.12\pm0.37$ &$17.89$&$5.09$&25.1(40) & $$\\
32&$1.82\pm0.03$&$37.2\pm3.6$&$8.91$&$2.49\pm0.05$&$8.77\pm0.36$ &$19.32$&$4.86$&27.2(40) & $2.69\pm0.05$\\
33&$1.72\pm0.03$&$31.0\pm3.9$&$5.86$&$2.50\pm0.05$&$8.36\pm0.37$ &$15.55$&$3.04$&28.2(40) & $2.58\pm0.05$\\
34&$1.80\pm0.02$&$43.6\pm3.8$&$9.78$&$2.59\pm0.05$&$8.97\pm0.37$ &$18.76$&$5.26$&25.6(40) & $2.60\pm0.06$\\
35&$1.76\pm0.02$&$49.3\pm3.7$&$10.15$&$2.58\pm0.05$&$8.75\pm0.36$&$17.71$&$5.36$&20.7(40) & $2.69\pm0.06$\\
36&$1.69\pm0.02$&$58.8\pm3.9$&$10.43$&$2.69\pm0.04$&$8.95\pm0.34$  &$17.43$&$5.33$&26.4(40) & $2.68\pm0.04$\\
37&$1.31\pm0.02$&$188.1\pm13.1$&$11.78$&$2.39\pm0.04$&$6.87\pm0.27$&$10.51$&$4.54$&27.6(40) & $2.12\pm0.20$\\
38&$1.30\pm0.03$&$147.8\pm14.7$&$9.04$&$2.36\pm0.04$&$6.94\pm0.31$ &$11.03$&$3.45$&23.8(40) & $2.40\pm0.06$\\
39&$1.30\pm0.02$&$167.8\pm13.7$&$10.35$&$2.31\pm0.03$&$6.97\pm0.27$&$10.61$&$3.97$&23.8(40) & $$\\
40&$1.30\pm0.02$&$171.3\pm13.1$&$10.71$&$2.35\pm0.03$&$6.94\pm0.27$&$10.61$&$4.12$&17.9(40) & $2.42\pm0.05$\\
41&$1.31\pm0.03$&$151.7\pm15.1$&$9.67$ &$2.30\pm0.04$&$6.64\pm0.30$&$10.57$&$3.74$&31.7(40) & $$\\
42&$1.31\pm0.02$&$133.8\pm12.7$&$8.43$&$2.26\pm0.03$&$6.51\pm0.28$ &$10.12$&$3.25$&18.2(40) & $2.42\pm0.04$\\
43&$1.33\pm0.02$&$169.2\pm12.2$&$11.55$&$2.30\pm0.04$&$6.77\pm0.27$&$10.35$&$4.57$&15.1(40) & $2.39\pm0.08$\\
44&$1.29\pm0.02$&$151.6\pm14.3$&$9.08$&$2.32\pm0.03$&$6.58\pm0.28$ &$9.85$ &$3.45$&23.8(40) & $2.43\pm0.07$\\
45&$1.59\pm0.02$&$70.5\pm3.9$  &$9.63$&$2.42\pm0.04$&$7.60\pm0.30$ &$12.17$&$4.63$&31.2(40) & $2.37\pm0.05$\\
46&$1.24\pm0.02$&$191.1\pm12.4$&$9.78$&$2.17\pm0.04$&$6.06\pm0.26$ &$6.15$ &$3.52$&23.1(40) & $2.20\pm0.16$\\
47&$1.20\pm0.02$&$174.5\pm12.2$&$7.90$&$2.05\pm0.05$&$5.74\pm0.28$ &$4.54$ &$2.73$&39.0(40) & $2.05\pm0.16$\\   
48&$1.21\pm0.02$&$112.1\pm7.8$ &$5.15$&$2.05\pm0.03$&$4.74\pm0.25$ &$3.53$ &$1.78$&28.6(40) & $2.19\pm0.06$\\
\hline
\end{tabular}
\par
\begin{list}{}{}
\item[${\rm ^{a}}$] -- normalization of the soft spectral component ${\rm N_{diskbb} = \left ( \frac{R_{in}(km)}{D_{10 kpc}} 
\right )^{2} cos(\theta)}$, where ${\rm R_{in}(km)}$ -- effective inner radius of the disk, ${\rm D_{10 kpc}}$ -- 
source distance in units of 10 kpc, $\theta$ -- inclination angle of the accretion disk\\
\item[${\rm ^{b}}$] -- bolometric flux of the soft spectral component in units of $\times 10^{-9}$ erg s$^{-1}$ cm$^{-2}$\\
${\rm ^{c}}$ -- equivalent hydrogen column density in units of $\times 10^{22} cm^{-2}$\\
\item[${\rm ^{d}}$] -- absorption corrected value of the total X-ray flux in the $3 - 20$ keV energy range in units of 
$\times 10^{-9}$ erg s$^{-1}$ cm$^{-2}$\\ 
\item[${\rm ^{e}}$] -- absorption corrected value of the soft spectral component flux in the $3 - 20$ keV energy range 
in units of $\times 10^{-9}$ erg s$^{-1}$ cm$^{-2}$\\
\item[${\rm ^{f}}$] -- photon index of hard spectral component derived from fitting of the HEXTE data.
\end{list} 
\end{table*}

\subsection{Timing analysis}

For the timing analysis the PCA {\em Binned, Single Binned} 
and {\em Event} mode data were used. We generated power density spectra 
(PDS) in the 0.001--256 Hz frequency range (2--13 keV energy band), 
combining the results of the summed Fourier transforms of a short 
stretches of data with $0.002$ s time bins for the $0.3 - 128$ Hz 
frequency range and a single Fourier transform on the data in $0.125$ s 
time bins for lower frequency band. The resulting spectra were 
logarithmically rebinned when necessary to reduce scatter at high 
frequencies and normalized to square root of fractional variability 
rms. White--noise level due to the Poissonian statistics corrected 
for the dead--time effects was subtracted \cite{vgch94,zhang95}.

To trace the evolution of the source timing properties, we fitted 
its power density spectra in the $0.01 - 100$ Hz frequency range 
to the analytic models using the $\chi^{2}$ 
minimization technique. For the approximation of the PDS obtained 
during the first 26 observations the combination of the band--limited 
noise (BLN) component (approximated by zero--centered Lorentzian 
function) and several quasi-periodic oscillations (QPO) 
(expressed by Lorentzian functions) was used. 
Beginning on the observation $\# 27$ (except for the 
observation $\# 33$) an additional power law component expressing 
the Very Low Frequency Noise (VLFN) component is required.

\begin{figure*}
\hbox{
\epsfxsize=9.0cm
\epsffile{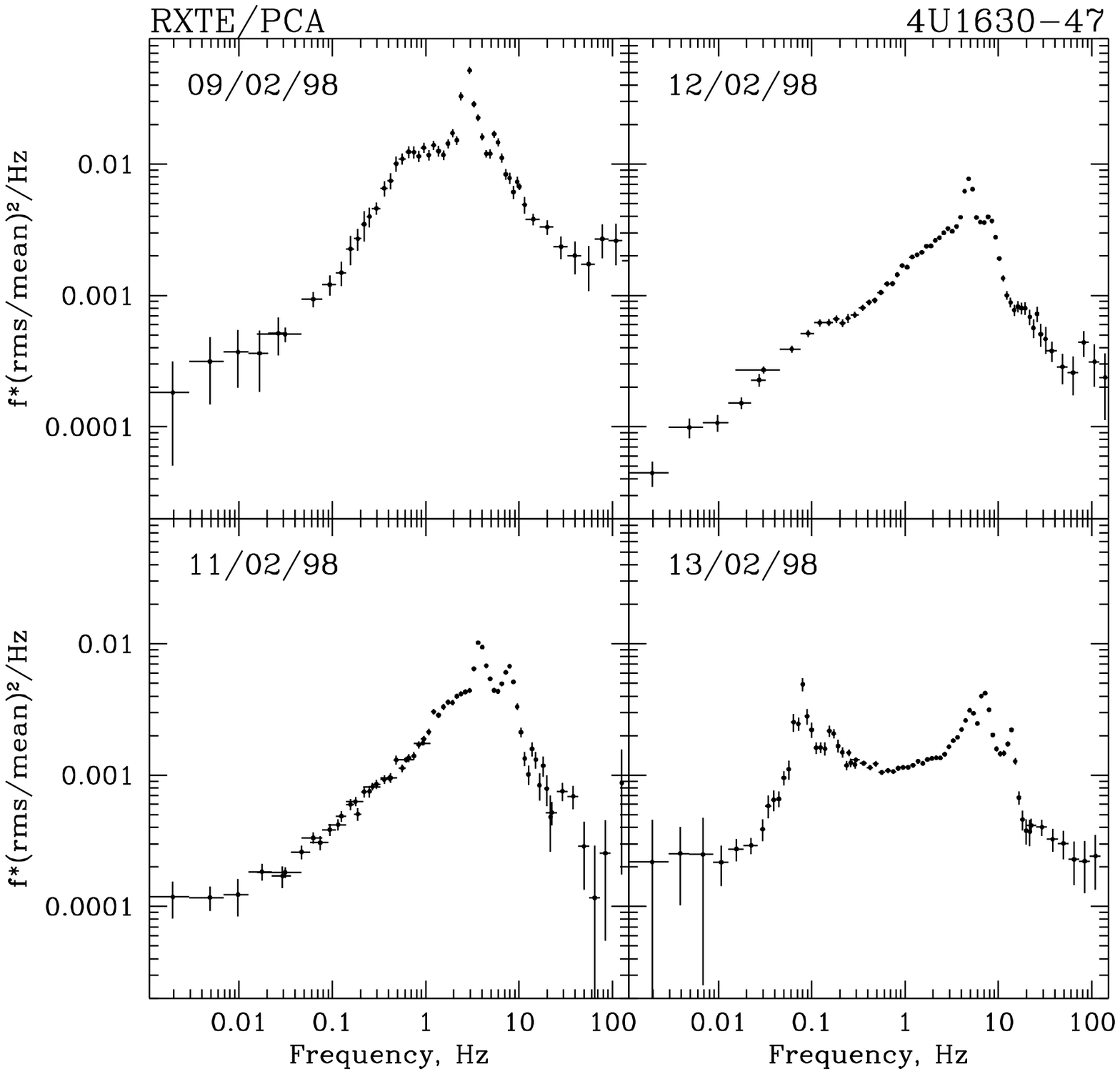}
\epsfxsize=9.0cm
\epsffile{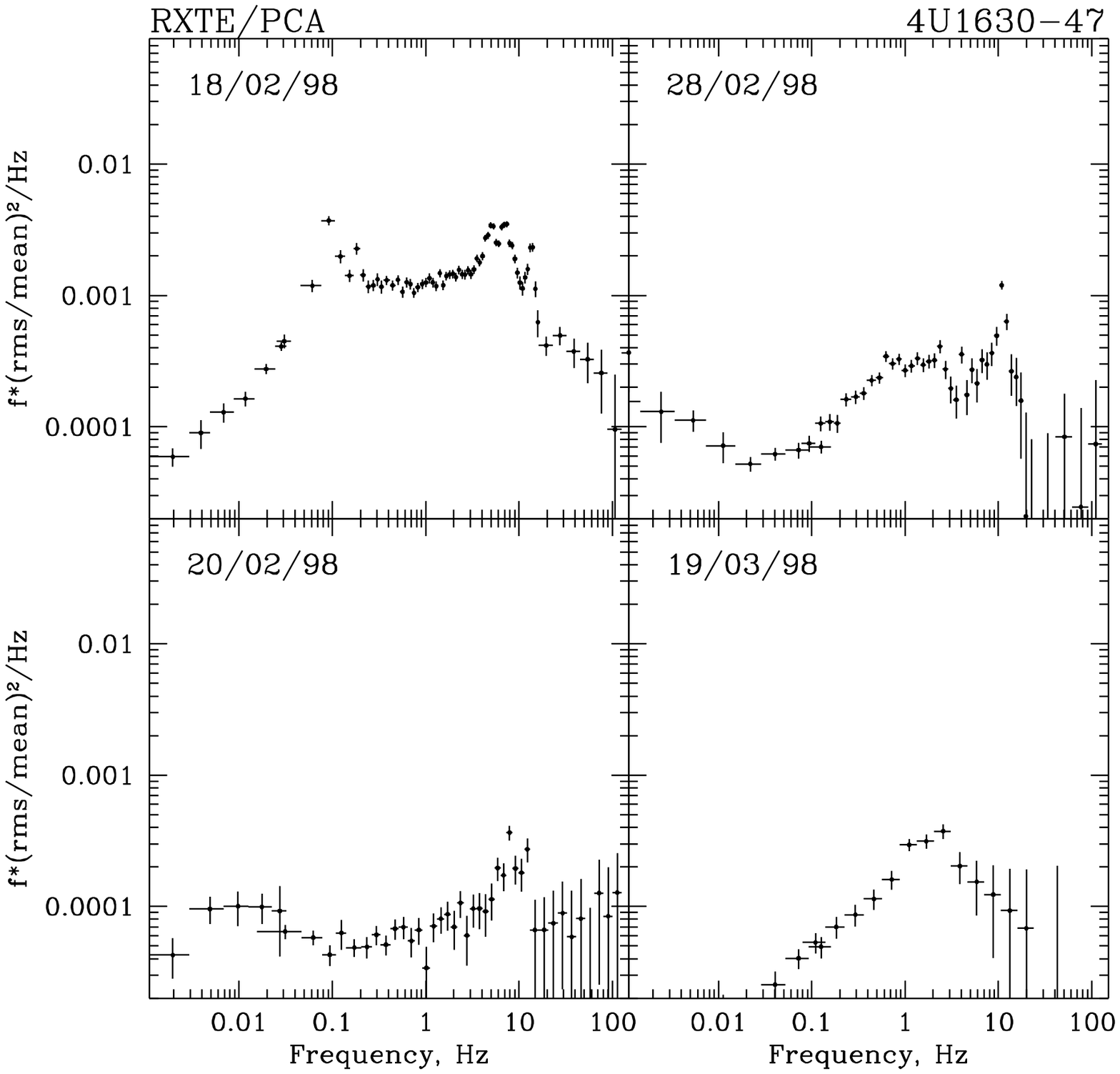}
}
\caption{Typical broad-band power density spectra of 4U1630-47 
in units of ${\rm f \times (rms/mean)^{2}}$/Hz during the initial 
rise and maximum phase of the 1998 outburst. $2 - 13$ keV 
energy band, PCA data. \label{pds}}
\end{figure*}

Typical examples of the broad--band power density spectra of 4U~1630--47 
during the 1998 outburst are shown in Fig. \ref{pds} (for the 
convenience, PDS are presented in units of ${\rm f \times (rms/mean)^{2}}$/Hz, 
representing the actual distribution of the variability amplitude over 
frequency). General parameters of the source power density spectra and 
their evolution during the outburst are presented in Table 
\ref{pds_params} and Figure \ref{timing_evol} respectively.

\begin{table}
\caption{The characteristics of the power density spectra of 4U~1630--47.
Parameter errors correspond to $1 \sigma$ confidence level. 
${\rm rms_{total}}$ represents total rms amplitude integrated over 
$0.02 - 100$ Hz frequency range, ${\rm f_{QPO}}$ and ${\rm rms_{QPO}}$ 
represent the centroid frequency and the rms amplitude of the Lorentzians 
used to approximate the QPO peaks.
\label{pds_params}}

\begin{tabular}{ccccc}
\hline
Date, UT &$\#$&${\rm rms_{total}, \%}$ &${\rm f_{QPO}}$, Hz&${\rm rms_{QPO}}$, $\%$ \\
\hline
09/02/98 &  (1)  & $27.97\pm0.24$ & $2.68\pm0.01$ & $14.84\pm0.90$\\
         &       &                & $5.60\pm0.07$ & $4.83\pm0.33$ \\
09/02/98 &  (2)  & $25.47\pm0.18$ & $3.21\pm0.01$ & $14.08\pm0.80$\\
         &       &                & $6.54\pm0.08$ & $3.93\pm0.30$ \\
11/02/98 & (3-5) & $12.34\pm0.09$ & $3.82\pm0.02$ & $4.85\pm0.30$ \\
         &       &                & $7.77\pm0.06$ & $4.79\pm0.30$ \\
12/02/98 & (6-10)& $10.83\pm0.05$ & $4.80\pm0.01$ & $4.44\pm0.19$ \\
         &       &                & $7.97\pm0.09$ & $3.80\pm0.35$ \\
13/02/98 &(11-13)& $10.71\pm0.05$ & $0.079\pm0.002^{1}$ & $4.19\pm0.49$\\
         &       &                & $4.74\pm0.06$ & $2.61\pm0.17$\\
         &       &                & $7.05\pm0.03$ & $3.62\pm0.11$\\
         &       &                & $13.59\pm0.08$& $1.99\pm0.14$\\
14/02/98 &(14-17)& $10.56\pm0.04$ & $0.094\pm0.002^{1}$ & $3.03\pm0.48$\\
         &       &                & $5.38\pm0.04$ & $4.53\pm0.16$\\
         &       &                & $7.75\pm0.13$ & $2.68\pm0.28$\\
         &       &                & $13.29\pm0.07$& $1.97\pm0.13$\\
15/02/98 & (18)  & $10.10\pm0.19$ & $0.067\pm0.005^{1}$ & $3.77\pm0.89$\\
         &       &                & $4.69\pm0.07$ & $2.07\pm0.29$\\
         &       &                & $7.07\pm0.04$ & $3.52\pm0.25$\\
         &       &                & $13.20\pm0.50$& $1.95\pm0.40$\\
16/02/98 &(19-21)& $9.89\pm0.10$  & $4.87\pm0.07$ & $2.56\pm0.22$\\
         &       &                & $6.94\pm0.06$ & $3.44\pm0.16$\\
         &       &                & $13.35\pm0.15$& $1.87\pm0.20$\\
17/02/98 &(22-23)& $10.57\pm0.18$ & $0.097\pm0.002^{1}$ & $3.28\pm0.84$\\
         &       &                & $5.23\pm0.09$ & $2.67\pm0.47$\\
         &       &                & $6.62\pm0.39$ & $4.69\pm1.04$\\
         &       &                & $13.83\pm0.34$& $1.51\pm0.50$\\
18/02/98 &(24-26)& $10.24\pm0.05$ & $0.091\pm0.002^{1}$ & $3.53\pm0.32$\\
         &       &                & $5.03\pm0.06$ & $3.23\pm0.29$\\
         &       &                & $6.89\pm0.15$ & $3.94\pm0.35$\\
         &       &                & $13.31\pm0.10$& $1.77\pm0.12$\\
19/02/98 &(27-28)& $2.13\pm0.23$  & $7.04\pm0.28$ & $1.32\pm0.26$\\
         &       &                & $12.56\pm0.85$& $1.43\pm0.30$\\
20/02/98 &(29-31)& $1.93\pm0.20$  & --            & --           \\
21/02/98 & (32)  & $2.74\pm0.37$  & --            & --           \\
21/02/98 & (33)  & $10.16\pm0.32$ & $0.052\pm0.010^{1}$ & $2.79\pm0.83$\\
         &       &                & $4.45\pm0.16$ & $1.88\pm0.40$\\
         &       &                & $7.04\pm0.06$ & $3.05\pm0.30$\\
         &       &                & $12.25\pm0.73$& $1.40\pm0.63$\\
22/02/98 & (34)  & $2.43\pm0.50$  & --            & --           \\
23/02/98 & (35)  & $3.09\pm0.16$  & --            & --           \\
24/02/98 & (36)  & $3.53\pm0.49$  & $9.50\pm1.65$ & $1.85\pm0.45$\\
25/02/98 & (37)  & $3.99\pm0.17$  & $10.93\pm0.83$& $1.75\pm0.43$\\
26/02/98 &(38-39)& $3.61\pm0.44$  & $11.19\pm0.43$& $1.86\pm0.34$\\
27/02/98 &(40-41)& $4.27\pm0.19$  & $10.33\pm0.44$& $1.78\pm0.30$\\
28/02/98 & (42)  & $3.74\pm0.27$  & $10.99\pm0.13$& $1.84\pm0.22$\\
01/03/98 & (43)  & $2.20\pm0.58$  & $10.76\pm0.59$& $1.77\pm0.47$\\
02/03/98 & (44)  & $1.98\pm0.69$  & $11.12\pm0.39$& $1.62\pm0.37$\\
09/03/98 & (45)  & $1.93\pm0.48$  & $7.87\pm0.26$ & $1.67\pm0.39$\\
12/03/98 & (46)  & $1.80\pm0.50$  & --            & --           \\
17/03/98 & (47)  & $1.96\pm0.75$  & --            & --           \\
19/03/98 & (48)  & $1.74\pm0.31$  & --            & --           \\
\hline
\hline

\end{tabular}
\par
\begin{list}{}{}
\item[$^{1}$] -- the QPO peak shows harmonic content
\end{list}

\end{table}

\section{LIGHT CURVE OF 4U~1630--47}

In four years RXTE satellite detected three outbursts from
4U~1630--47 (Fig. \ref{lc_general}).  The beginning of 1998
outburst was predicted by Kuulkers et al. (1997) with an
accuracy of few days.  This prediction was based on the time 
of the 1996 outburst, and the mean period over the epoch 1987--1996.
However, the outburst of 1999 occurred
much earlier than was expected from both the interval
of $\sim$ 690 days suggested by Kuulkers et al. (1997), 
and from earlier calculated recurrence
interval 602 $\pm$ 3 days \cite{parmar95}.
Because the recurrence time of the outbursts appears not to be
absolutely stable, we suggest that this time is not linked to
the parameters of binary orbit, but instead is due to the typical 
time of mass collection in the accretion disk.  Recurrent
outbursts with typical time 440 $\pm$ 30 days were observed 
from another Galactic black hole candidate GX~339--4 in 1991--1994
\cite{harmon94}, but later the recurrence was broken.  
In case of GX~339--4 the orbital period was measured to be 
14.8 h \cite{cal92}.

The three outbursts differ significantly in their shape. While
the outbursts of 1996 and 1999 years remind low/high state transitions
in persistent X-ray sources as Cyg X-1 \cite{cui98} and GX339-4 
\cite{bel99}, 1998 outburst was of fast-rise-exponential -decay 
(FRED) type typical for many X-ray Novae \cite{chen97}.

The light curve of the 1998 outburst presented in more detail in 
Fig. \ref{lightcurve}. The flux from 4U1630--47 rose for about 5 days 
before reaching the level of about 200 mCrab.  The maximum of the outburst 
was composed of three plateaus with almost constant flux at each plateau 
and fast jumps between them. The second plateau was the highest one with 
maximum flux above 400 mCrab. The duration of each plateau was 6--7 days. 
A month after the beginning of the outburst quasi-exponential decline 
started, with secondary maxima, as typical for many X-ray transients.

\section{EVOLUTION OF THE SOURCE DURING THE OUTBURST OF 4U~1630--47}
During 1998 outburst the source demonstrated a variety of different
spectral and timing states. Thanks to good coverage of the outburst
by RXTE pointing observations we had an opportunity to study the 
changes in detail.  A summary of the states observed is 
presented in Table \ref{states}.

\begin{table*}
\caption{Summary of the source behavior for different states.
\label{states}}
\begin{tabular}{cccccccc}
\hline
State & when observed & $F_{total}^a$ & PL slope$^b$ & $E_{cut}^c$, keV & $f_{soft}^d$ & $rms_{total}, \%$ & QPO\\
\hline
Rise phase & Feb 9--11, 98 & 4--12 & 1.7--2.2 & 60--70 & $<$ 20 $\%$ & 12--28 & 2.5--4 Hz (5--15$\%$)\\
\hline
Very high & Feb 12--18 ({\it Plateau 1}) & 12--15 & 2.2--2.5 & $>$200 & 14--22 $\%$ & 10--11 & 4 QPOs in 0.06--14 Hz\\
          & Feb 19--24 ({\it Plateau 2}) & 17--20 & 2.5--2.7 & $>$200 & 23--30 $\%$ & 2--3.5 & weak or no\\
          & Feb 25--Mar 2 ({\it Plateau 3}) & 10--11 & 2.3--2.4 & $>$200 & 30--45 $\%$ & 2--4 & 10-11 Hz $(<2\%)$\\
\hline
High$^e$ & Mar 12 -- May 16 & $\leq$ 6 & 2.0--2.5 & $>$200 & $>$ 50 $\%$ & $<$4 & no or weak\\
\hline
Low$^e$ & after May 18, 98 & $\leq$ 1 & 1.5--2.0 & 50--100 & $<$ 10 $\%$ & $>$20 & $\sim$0.1--1 Hz\\
\hline
\end{tabular}
\par
\begin{list}{}{}
\item[$^{a}$] -- absorption corrected value of the total X-ray flux in the $3 - 20$ keV energy range in units of $\times 10^{-9}$ erg s$^{-1}$ cm$^{-2}$\\
\item[$^{b}$] -- index of power-law approximation of hard component\\
\item[$^{c}$] -- high-energy cut-off\\
\item[$^{d}$] -- fraction of the soft spectral component flux in 
the $3 - 20$ keV total flux\\
\item[$^{e}$] -- only few typical observations for these states were analyzed\\
\end{list} 
\end{table*}

\subsection{The rise of the outburst}
RXTE pointed observations started at Feb 9, 1998, a week after the 
beginning of the outburst detected by ASM/RXTE. Several PCA and HEXTE 
observations were carried out when the source was still in the rise phase. 
The broad--band energy spectrum of 4U1630-47 during these observations 
($\#\# 1 - 5$) can be generally described by sum of a strong hard component 
approximated by a power law model with a photon index $\sim 1.8 - 2.2$ and 
a high energy cut--off at $\sim 60 - 70$ keV and a weak soft spectral 
component with color temperature $\sim 1.7 - 1.8$ keV contributing 
$< 20 \%$ to the total source flux in the $3 - 20$ keV energy band 
(Fig. \ref{spectra}, \ref{spec_evol}). The monotonic rise of the source 
X-ray flux from $\sim 4 \times 10^{-9}$ to $\sim 1.2 \times 10^{-8}$ 
erg/s/cm$^{2}$ \footnote{ -- absorption--corrected value of the source 
X-ray flux in the $3 - 20$ keV energy band} during this period was 
accompanied by gradual steepening of the hard spectral component and rise 
of the soft component flux (Fig. \ref{spec_evol}, Table \ref{params}).  
   
The broad-band power density spectrum (PDS) of the source is dominated by 
strong band-limited noise (BLN) component and a complex of a relatively 
narrow QPO peaks 
(${\rm \Delta f/f \sim 0.1 - 0.3}$, where ${\rm \Delta f}$ and 
${\rm f}$ are the width and the centroid frequency of the QPO peak) 
placed near the breakpoint in the slope of the BLN continuum (Fig. \ref{pds}). 
For the first two observations (Feb. 9, 1998) the centroid frequencies of 
the QPO peaks are harmonically related, while for further observations 
this relation breaks. The evolution of the temporal properties of the 
source during this phase of the outburst is characterized by monotonic 
decrease of the total fractional rms amplitude of the rapid aperiodic 
variability from $\sim 28 \%$ to $\sim 11 \%$, accompanied by increase 
of the BLN break and QPO centroid frequencies (i.e. the systematic shift 
of the effective maximum of the power spectral density distribution to the 
higher frequencies; Table \ref{pds_params}).

The complex of the X-ray properties of 4U1630-47 during this stage of the 
outbursts is very similar to that of some Galactic black hole candidates 
observed during their high--to--low and low--to--high state 
transitions \cite{tl95}. In fact, there are several general properties 
of the source showing that this state of the source does not match neither 
'canonical' low nor high/very--high and probably corresponds to 
the transition between these states (see discussion below).

\subsection{Very high state (VHS)}

The peak of the 1998 outburst lasted for $\sim 25 - 30$ days and 
followed by an exponential decay. The peak itself can be roughly 
divided into three plateau at different flux levels, distinguished  
by the quantitative differences in the source spectral and temporal 
behavior. In spite of all these differences, the general complex 
of X-ray properties of the source during the whole period of peak 
flux allows to classify it as a 'canonical' very--high state 
established for Galactic black hole candidates \cite{miya91,miya93,tl95}.

{\it Plateau 1}. Total X-ray flux from the source detected with PCA 
rose slowly for observations $\#\# 6 - 26$ with average level of $\sim 
1.4 \times 10^{-8}$ erg/s/cm$^{2}$ (Fig \ref{spec_evol}). The broad--band 
energy spectrum in this state can be satisfactory described by the sum 
of two components: a relatively weak soft component (contributing 
$< 20 \%$ to the total X-ray flux in the $3 - 20$ keV energy band) with 
color color temperature $\sim 1.6 - 1.7$ keV and a strong dominating hard 
component which has a power law form without high energy cut--off up 
to $\sim 150$ keV (Fig. \ref{spectra}).

\begin{figure}
\epsfxsize=8.5cm
\epsffile{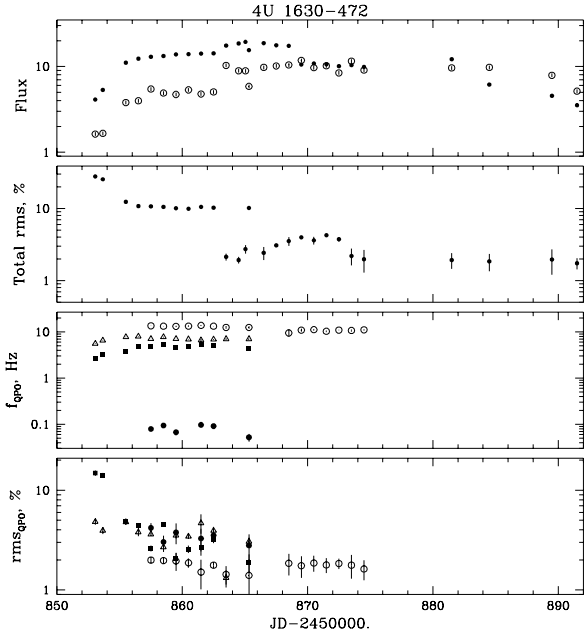}
\caption{The evolution of main parameters of the source power density 
spectrum (three lower panels) along with the X-ray flux histories 
(upper panel). Filled and hollow circles in upper panel correspond to the 
total X-ray flux in the $3 - 20$ keV energy range and the bolometric flux 
of the soft thermal component. 'Total rms' denotes the integrated fractional 
rms in the $0.01 - 128$ Hz frequency range. Bottom panels show the evolution 
of the centroid frequency and integrated rms of various QPO components of 
the PDS.  
\label{timing_evol}}
\end{figure}

The character of the X-ray variability of 4U~1630-47 during this period 
is of special interest. As it is clearly seen from the upper panel 
of Fig. \ref{dip_lc_similar}, the light curve of the source is marked by 
presence of quasi-regular modulations with period of $\sim 10 - 20$ s. 
Similar variability was observed in the Galactic microquasars 
GRS 1915+105 \cite{Morgan97} and GRO J1655-40 \cite{rem99}. See 
Fig. \ref{dip_lc_similar} and discussion in \S5.2. 

The broad--band PDS of 4U~1630-47 is strongly dominated by BLN component, 
total amplitude of the source rapid aperiodic variability in the $0.01 - 
100$ Hz frequency band was at the approximately constant level of $\sim 10 \%$ 
(Fig. \ref{pds}, Feb. 13 and 18 observations, Table \ref{pds_params}). 
Two groups of QPO peaks are a generic feature of the source power density 
spectrum (Fig. \ref{pds}, Table \ref{pds_params}): first group located in 
the $\sim 0.05 - 0.2$ Hz frequency range, caused by aforementioned 
quasi--regular dipping and second group of QPO peaks at $\sim 5$, $\sim 7$ 
and $\sim 13$ Hz. Contrary to the highly variable QPO features at $\sim 5$ 
and $\sim 7$ Hz, the 13 Hz QPO peak demonstrates relatively high stability 
of the centroid frequency and integrated rms amplitude (Table 
\ref{pds_params}).   

{\it Plateau 2}. Beginning on the Feb. 19 the flux from the source 
jumped up significantly and stayed high and variable until the 
observation $\# 36$ (Fig. \ref{lightcurve}). During this period the 
high energy part of the source spectrum became steeper, while the 
color temperature of the soft thermal component slightly rose to 
$\sim 1.8$ keV. Soft spectral component contribution 
was still below $\sim$ 30\% (Fig.\ \ref{spectra},\ \ref{spec_evol}; Table 
\ref{params}), and its fast variability - significantly weaker 
(total rms variability $< 3 \%$). QPOs became hardly detectable 
(Fig. \ref{pds}, Feb. 20 observation). A notable exception was the 
observation $\# 33$, when source flux and behavior returned to the 
pattern of {\it Plateau 1}. 

{\it Plateau 3}. Since the observation $\# 36$ (Feb. 24) the X-ray flux of 
4U~1630-47 dropped to $\sim 1.1 \times 10^{-8}$ erg/s/cm$^{2}$ level and 
remained rather stable until the observation $\# 44$ (March 2). The 
quantitative characteristics of the broad--band energy spectra are quite 
different from that of the previous state: color temperature of the soft 
spectral component decreased to $\sim 1.3$ keV, the contribution of the 
soft component to the total X-ray luminosity increased up to the $\sim 40 \%$ 
-- level, while the high energy part of the spectrum significantly 
flattened (average photon index, $\alpha \sim 2.3$). The fractional 
variability of 4U~1630-47 somewhat increased (rms $\sim 4 \%$).
The PDS is dominated by BLN component with a peak at $\sim 1 - 2$ Hz, 
and, most notably, a prominent QPO feature with the frequency of 
$\sim 10 - 11$ Hz and rms amplitude $\sim 2 \%$ showed up
(Fig. \ref{pds}, \ref{timing_evol}; Table \ref{pds_params}). 
It should be noted, that this QPO feature again demonstrates relatively 
high stability of the centroid frequency and rms amplitude which 
allows us to treat it as a successor of the same nature as 
13 Hz--QPO observed during the {\it Plateau 1}. 

\subsection{High state (HS)}

After {\it Plateau 3} the flux from the source started to decline 
quasi--exponentially. Behavior of the source during the decline 
(after 19/03/1998 (observation $\# 46$) may be well described by 
standard high state of black hole candidates (see \cite{tl95} for 
a review and references wherein). 

\subsection{Low state (LS)}

At late stage of 1998 outburst the source underwent the transition
from high to low state, which is also very typical for all Galactic 
black hole candidates and for X-ray Novae in particular.  
Hard power law spectrum
was observed, in combination with rather strong fast variability at
the level of $\sim 5 - 10 \%$, and with QPO of variable frequency in
the range $\sim 0.1 - 3$ Hz.

\section{DISCUSSION}

\subsection{Spectral and timing properties of 4U~1630-47 during the rise phase 
of the outburst}

\begin{figure}
\epsfxsize=8.5cm
\epsffile{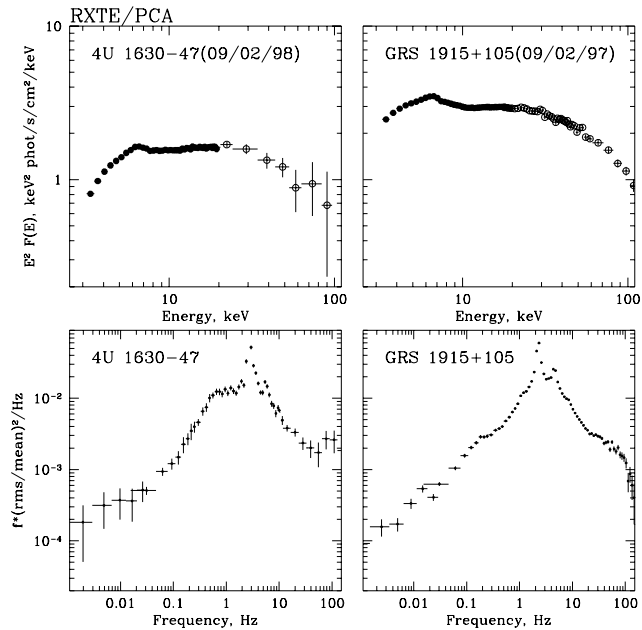}
\caption{Demonstration of the similarity at the rise phase
between 4U~1630--47 ({\it left} panels) and GRS~1915+105 ({\it right} panels). 
Energy spectra are presented in {\it upper} panels, and power density 
spectra -- in {\it lower} panels.
\label{rise_phase}}
\end{figure}

We examined the 4U~1630-47 observations during the initial rise 
of 1998 outburst, and tried to understand how this 
state fits standard scheme of the 'canonical' states established for the 
black hole candidates \cite{tl95}. The broad--band energy and power 
density spectra of the source during this stage of the outburst are 
shown in Fig.\ref{rise_phase}. 

The energy spectrum is relatively hard ($\alpha \sim 2$) and has 
clear high--energy cut--off at the energies $\sim 60 - 70$ keV, the 
contribution of the soft thermal component to the total X-ray luminosity 
is small $< 20 \%$. The same type of energy spectrum was previously 
observed during the initial rise phase of the outbursts of other X-ray 
Novae GS/GRS 1124-68 (Nova Muscae 1991; \cite{ebi94,tak97}) and 
KS/GRS~1730--312 \cite{bor95,trud96}. 

There is also a striking resemblance between the X-ray 
properties of 4U~1630--47 during the rise phase of the 1998 outburst 
and GRS~1915+105 in the low luminosity state \cite{Trudolyubov99.1}. 
Corresponding broad--band energy and power density spectra of these 
sources are shown in Fig.\ref{rise_phase}.

The spectrum and fast variability of 4U~1630-47 during this period is 
very similar to several Galactic black hole candidates, namely, 
Cyg X-1 \cite{bel96,cui97}, GX~339-4 \cite{mvdk96}, GS~1124-68 
\cite{miya94,tak97}, GRO~J1655--44 \cite{mbk98}, observed in 
the so-called intermediate state during their high--to--low 
and low--to--high state transitions. This state comprises properties 
of both low and very--high states and evidently corresponds to the 
transition between them. The X-ray flux of the 
source during the rise phase is significantly higher than in the typical 
low luminosity state $< 10^{-9}$ erg/s/cm$^{2}$. At the end of 
rise phase the flux is approaching the level very--high state.
As in low state, the spectrum is hard, furthermore, it 
exhibits high-energy cut-off, which is typical for low state,
but not for very--high state.

While the energy spectrum resembles low state, the power density 
spectrum differs from the typical PDS in the low state: the frequency 
of the characteristic peak of the BLN continuum ($\sim 3 - 10$ Hz) is an 
order of magnitude higher than usual for Galactic black hole candidates in 
LS, but both general shape and characteristic frequencies of the PDS are 
consistent with that of the VHS. 

The evolution of the spectral and timing parameters of 4U~1630-47 during 
the rise phase of the outburst can be generally understood in the framework 
of the two-phase model of the accretion flow around the compact object 
(i.e. the composition of a hot inner comptonization region \cite{st80} 
and surrounding optically thick accretion disk \cite{ss73}). The 
interaction between these two distinct regions determines the properties 
of the spectrum and variability of the source. Assuming the QPO phenomenon 
to be related to the dynamical time scale on the boundary between the hot 
inner region and the outer accretion disk \cite{msc96,tlm98}, 
we can treat the observed increase of the QPO centroid frequency as an 
indication of the inward motion of this boundary during the rise phase 
of the outburst. This interpretation is supported by simultaneous shift 
of the maximum of the PDS band-limited noise component (depending on the 
characteristic radius of the inner comptonization region) and softening 
of the energy spectrum. Aforementioned similarity between the rise of 
4U~1630-47 1998 outburst and the outbursts of other X-ray Novae, allows 
us to suggest that in all these cases the approach of the inner edge of 
an optically thick accretion disk to the black hole causes the rise of the 
soft X-ray emission and softening of the broad-band energy spectrum. 
Alternatively, the hardening of the energy spectrum and decrease of the 
QPO centroid frequency and frequency of the BLN component maximum observed 
during the final stages of the X-ray Novae outbursts \cite{miya94,mbk98}, 
could be explained in terms of the outward motion of the boundary between 
the inner region and an optically thick accretion disk.

\begin{figure}
\epsfxsize=8.5cm
\epsffile{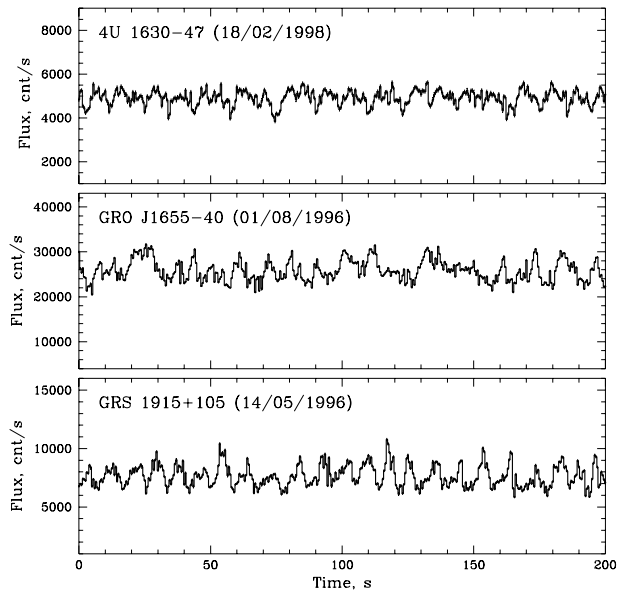}
\caption{Similar X-ray variability patterns for 4U1630-47 (upper panel), 
GRO~J1655--40 (middle panel) and GRS~1915+105 (lower panel) according to 
the RXTE/PCA observations ($2 - 30$ keV energy range, fluxes 
correspond to the 5 Proportional Counter Units). \label{dip_lc_similar}}
\end{figure}

\begin{figure*}
\epsfxsize=17.0cm
\epsffile{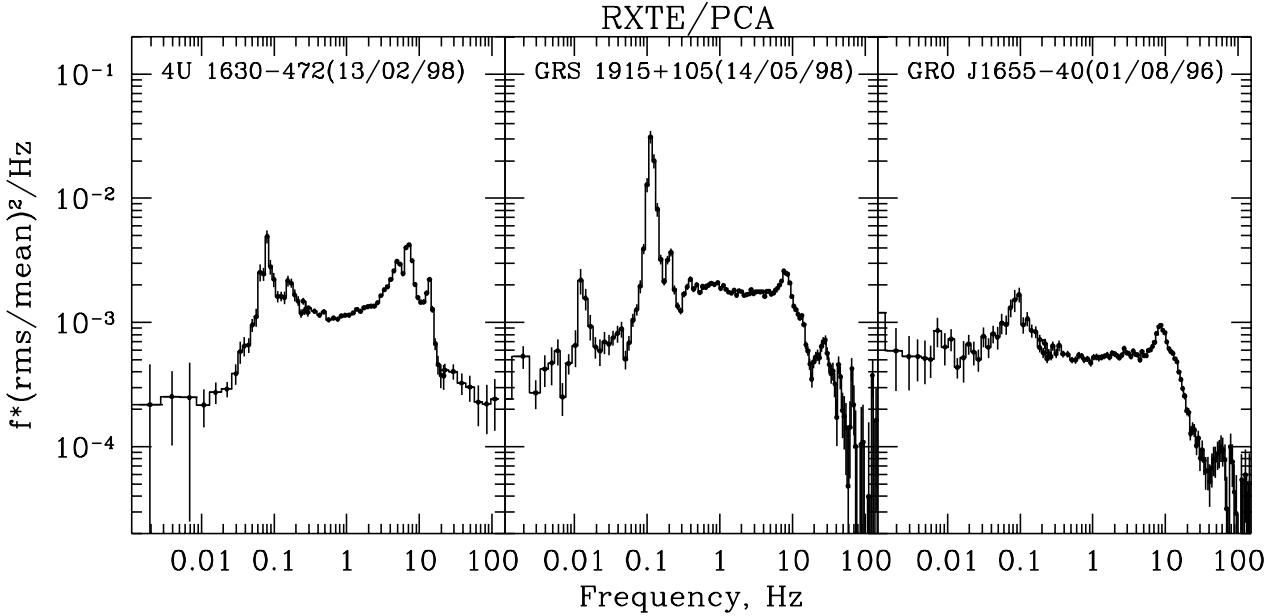}
\caption{Power density spectra of 4U~1630--47, GRO~J1655--40 and GRS~1915+105 
for the observations with similar slow variability pattern (Fig. \ref{dip_lc_similar}) PCA data. \label{dip_pds_all}}
\end{figure*}

\subsection{Quasi-regular variability in 4U~1630--47 on time scales of 
tens of seconds}

\begin{figure}
\epsfxsize=8.5cm
\epsffile{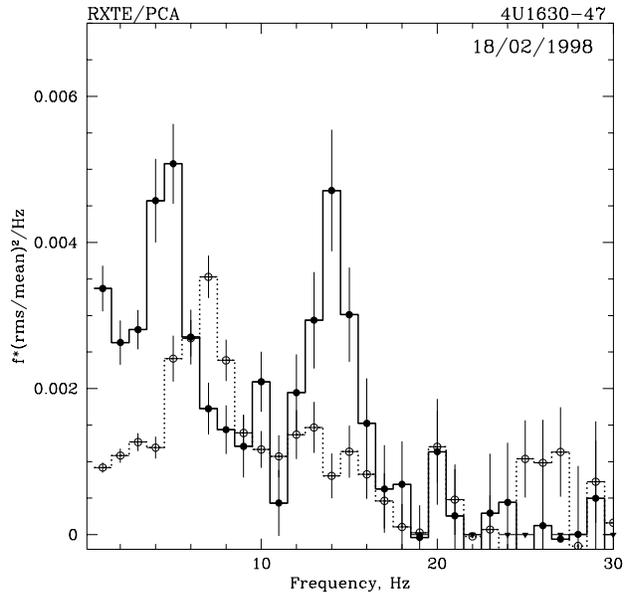}
\caption{Comparison of the source power density spectra in units of 
${\rm frequency \times (rms/mean)^{2}}$/Hz during the low (filled circles, 
solid lines) and high (hollow circles, dotted lines) flux episodes for 
the same observation on {\it Plateau 1} (observation $\# 24$, $2 -- 13$ keV). 
QPO at $\sim$ 13 Hz is prominent for low fluxes, and disappears for
high fluxes. This may be attributed to the breakthrough of the shock, or
to inward motion of the accretion disk (see the text).
\label{4u_high_low_pds}}
\end{figure}

The variability of spectral and timing properties of the source during 
the {\it Plateau 1} is of special interest. As it is clearly seen from 
Figure \ref{dip_lc_similar} (upper panel), the light curve of the 
source is marked by presence of quasi-regular modulations with period 
$\sim 10 - 20$ s. This type of variability is seen in the PDS as a QPO 
peak in the range $0.05 - 0.1$ Hz showing harmonic content (see Table 
\ref{pds_params}). We have performed detailed analysis of the variability 
of spectral and fast timing parameters of the source associated with this 
type of flux variability.  All observations in {\it Plateau 1} were used
for this analysis.  We extracted energy and power 
density spectra of the source separated according to the level of the 
source energy flux using the data in the {\em 'Binned'} and {\em 'Event'} 
modes. The data were segregated according to the total count rate 
averaged over 2s time intervals. The range from minimum to maximum flux
was divided into eleven equal parts. PDS corresponding to each luminosity 
interval was fitted to the analytical model consisting of a band--limited component and up to three Lorentzians. Energy 
spectra were approximated by a multicolor disk plus power law model with correction to the low energy absorption. Energy spectra and PDS for the highest and the lowest flux intervals are presented in Fig. \ref{4u_high_low_spec_rat}, \ref{4u_high_low_pds}.

There is a significant difference between the energy spectra of 
the source during the periods of the low and high flux: the rise of the 
X-ray flux is accompanied by softening of the hard spectral component and 
increasing of the effective temperature of the soft spectral component 
(Table \ref{tab_low_high}, Fig. \ref{4u_high_low_spec_rat}).

\begin{table}
\caption{Main spectral parameters of 4U~1630--47 corresponding to the low 
and high flux levels, derived using the combination of the multicolor disc 
blackbody and power law models with correction to the interstellar 
absorption, fixed at the value of $8 \times 10^{22}$ cm$^{-2}$. (observation 
$\#$ 24 (18/02/1998), PCA data, $3 - 20$ keV energy range). Parameter errors 
correspond to $1 \sigma$ confidence level for the assumed $1 \%$ systematic 
uncertainty of data. \label{tab_low_high}}
\begin{tabular}{cccccc}
\hline
      & $T_{d}$, keV & $\alpha_{pl}$ & $f_{total}^{a}$ & $f_{soft}^{b}$ & $\chi^{2}$ \\
\hline
Low   & $1.65\pm0.06$& $2.27\pm0.11$ & $1.20$ & $0.29$ & $34.8(21)$\\
High  & $1.91\pm0.20$& $2.52\pm0.13$ & $1.62$ & $0.35$ & $29.2(21)$\\
\hline
\end{tabular}
\par
\begin{list}{}{}
\item[$^{a}$] -- absorption corrected value of the total X-ray flux in the $3 - 20$ 
keV energy range in units of $\times 10^{-9}$ erg s$^{-1}$ cm$^{-2}$\\
\item[$^{b}$] -- absorption corrected value of the soft spectral component flux in 
the $3 - 20$ keV energy range in units of $\times 10^{-9}$ erg s$^{-1}$ cm$^{-2}$
\end{list}
\end{table}

\begin{figure}
\epsfxsize=8.5cm
\epsffile{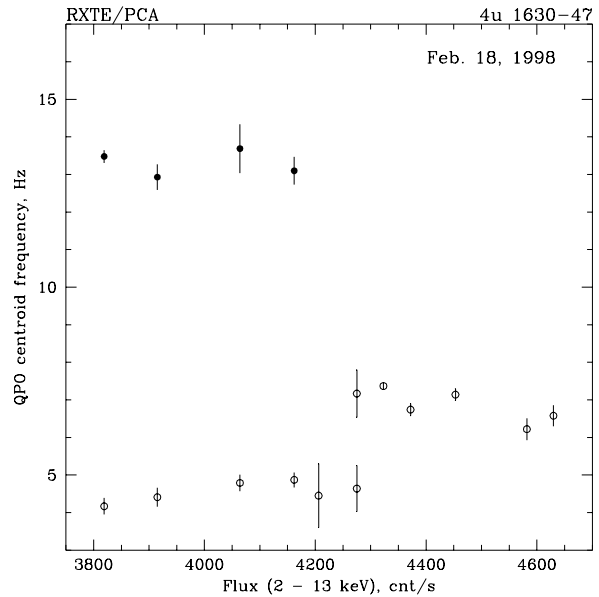}
\caption{Dependence of the QPO centroid frequencies on the source flux in 
the $2 - 13$ keV energy band (5 PCUs of PCA detector) for the Feb. 18 observation ($\# 24$). The $\sim 13$--Hz QPO feature is not detectable for fluxes above $\sim 4200$ cnt/s. The centroid of the second QPO feature is jump--shifted to higher frequencies at the same flux level. These data may be interpreted as an existence of two distinct quasi-stable states with sharp transition between them. \label{qpo_lumin}}
\end{figure}

The properties of the source power density spectra corresponding to the 
periods of high and low flux are quite different. Low flux PDS has 
relatively high level of the total rms variability amplitude 
of $\sim 6.7 \%$ ($1 - 128$ Hz frequency band) and is characterized by 
the presence of two prominent QPO features at $\sim 4.5$ Hz and $\sim 13$ Hz 
(Fig. \ref{4u_high_low_pds}, \ref{qpo_lumin}). The increase of the source flux is accompanied by drastic changes in the properties of the PDS: the  
$13$-Hz QPO feature disappears, while the lower frequency QPO peak 
shifts to $\sim 6 - 7$ Hz(Fig. \ref{4u_high_low_pds}, \ref{qpo_lumin}). In addition, high flux PDS has much lower total rms amplitude of $\sim 3.5 \%$ 
($1 - 128$ Hz frequency band). 

It is notable that somewhat similar type of variability was observed in 
Galactic microquasars GRS~1915+105 and GRO~J1655-40 \cite{Morgan97,rem99}; 
see Fig. \ref{dip_lc_similar}, \ref{dip_pds_all}. 
We speculate that this mode might be common for black hole binaries 
emitting at certain luminosity level. It is remarkable that in all 
cases, this mode corresponds to a relatively narrow interval of source 
luminosities. For 4U~1630--47 this type of variability was observed 
only for the {\em Plateau 1} observations, disappearing during 
the neighboring {\em Plateau 2} and {\em Plateau 3}. 
Moreover, when in the middle of the 
{\em Plateau 2} the source flux dropped to the level of {\em Plateau 1}, 
the pattern of variability immediately returned back (Table 
\ref{pds_params}, observation \#33).

\subsection{Correlation between the spectral and timing parameters}

Our analysis demonstrates a strong correlation between the  
energy spectrum of the source and its fast variability. During the 
initial rise of the outburst, and subsequent very--high state, the 
rise of the flux is accompanied by an increase of the QPO frequency, and
decrease of the fast variability. For each plateau a
stable flux level corresponds to stable spectral and 
timing parameters (Fig. \ref{spec_evol}, \ref{timing_evol}). It should 
be noted, that the complex of spectral and timing properties of 4U~1630--47 
is very sensitive to the source luminosity level: changes of the 
source flux are accompanied by the changes in the energy 
spectrum and PDS. As an example, observation $\#$ 33 in the middle
of {\it Plateau 2} has a flux typical of {\it Plateau 1}, and also
{\it Plateau 1} spectral and temporal behavior.

We note the correlation of the total source fractional variability with the 
flux of the hard spectral component and with the total X-ray flux integrated 
over $2 - 30$ keV band during the initial rise and maximum phase of the 
outburst ({\it Plateaus 1 and 2}). A similar correlation was first observed 
for GS/GRS~1124--68 (Nova Muscae 1991; \cite{miya94}) and later for 
many other Galactic black hole candidates.

\begin{figure}
\epsfxsize=8.5cm
\epsffile{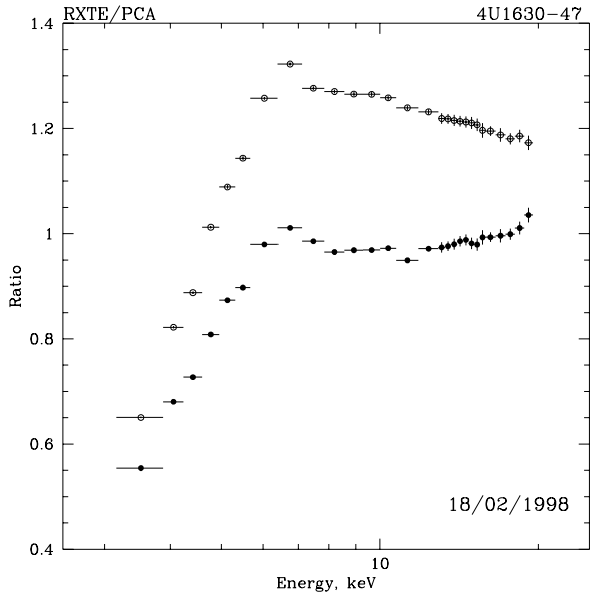}
\caption{The changes in 4U1630-47 energy spectrum between the low 
(filled circles) and high (open circles) flux episodes for the
same observation on {\it Plateau 1} (observation $\# 24$). 
Ratios of energy spectra to a power law 
with photon index $\alpha = 2.5$ are shown. 
The differences in both the temperature of soft component and the 
slope of hard component (also Table \ref{tab_low_high}) 
can be interpreted as an inward motion of 
the accretion disk inner edge during the periods of higher flux.
\label{4u_high_low_spec_rat}}
\end{figure}

The combination of the hot, optically thin quasi-spherical inner corona
surrounded by outer optically thick accretion disk is considered 
nowadays as a plausible geometry for accretion onto a black hole 
(see e.g. \cite{dove97}). A shock front formation between these two parts 
of the accretion flow is recognized as essential component of the picture 
in some models \cite{ct95}. We believe, that such models are able to 
reproduce qualitatively the correlations of spectral and timing properties 
observed for 4U~1630-47.  

It is often suggested that the QPO phenomenon is caused by interaction 
between the two aforementioned distinct parts of the accretion flow, 
and occurs on the dynamical time scale at the boundary of these 
regions \cite{msc96,tlm98}. Therefore, the change of the QPO centroid 
frequency is readily interpreted as an indication of the change in the 
effective radius of the boundary, which in turn is directly linked to 
the luminosities of the soft and hard components of the energy 
spectrum, and with the overall shape of energy spectrum.

Let us discuss the correlated fast and slow variability and spectral 
evolution, observed in 4U~1630-47 during the {\em Plateau 1} (see section 
\S5.2), in the framework of this model. The observation of 13-Hz QPO 
only at lower fluxes shows that the structure associated with this QPO 
is present when flux level is low, and disappears or substantially weakens 
for higher fluxes. It is plausible that this structure might 
be a shock \cite{ct95,msc96}. Then the 13-Hz QPO could be 
caused by resonance oscillations of the shock, while the slow 
oscillations on the time scale of $10 - 20$ s would correspond to the 
time of matter accumulation at the shock front (i.e. the shock 
stability time scale). A sudden breakthrough of the matter through the shock, 
or inward drift of the shock front leads to the disappearance 
of the 13-Hz QPO in PDS, or to the increase of QPO centroid frequency 
with simultaneous decrease of the its strength below the threshold 
of our detection. The accompanying inward drift of the inner boundary 
of the accretion disk causes the increase of the centroid frequency
of the second ($4 - 8$ Hz) QPO peak in PDS, and the increase of the soft 
component flux in agreement with observations 
(Fig. \ref{4u_high_low_spec_rat}). The softening of the high energy part 
of the spectrum is probably caused by the increase of Thomson 
optical depth and decrease of the temperature of a hot corona behind the 
shock front. In fact, the accretion inside the shock in this case
may be described also by a bulk motion Comptonization model \cite{ct95,tmk97}. 
On the other hand, even in the absence of a shock, the source behavior 
could be explained by the movement of the boundary between the inner hot 
region and optically thick accretion disk caused by some kind of 
accretion disk instability \cite{tlm98,Trudolyubov99.2}.

The variations in the characteristic QPO frequencies hint at a
change of some typical radius of the system, in particular, the inner 
radius of the optically thick accretion disk. Because the luminosity 
of the disk depends on both the accretion rate and the inner radius of 
the disk \cite{ss73}, we can not simply attribute any changes in 
the source luminosity to the change of the accretion rate onto the 
central object. In fact, the luminosity can vary significantly for 
the same accretion rate, if the geometry of the system and relative 
contribution from disk and corona component change. The observation 
of several plateaus at the peak of the outburst shows the 
existence of some quasi-stationary modes of the accretion for
this state.  The high state, in contrast, exhibits
a monotonic and quasi-exponential decrease, with associated
spectral and temporal behavior changes.

\subsection{Comparison with other outbursts}
Of the three outbursts of 4U~1630-47 observed with the
RXTE the outbursts of 1996 and 1999 look most similar.
In both cases the flux from the source in ASM/RXTE energy band
(1.3 --12 keV) rose fast, was high and chaotically variable
for some time, then faded quickly. The spectra at maximum
were typical for the high state of Galactic black holes. The transition 
from high to low state was detected by the PCA and HEXTE during 
the 1999 outburst \cite{mccol99}. It is plausible that a similar 
transition would have been observed for the outburst of 1996, if pointed
observations were available. Overall behavior
of the source in these outbursts is that of a low-high-low state
transitions, which is typical for "persistent" Galactic black hole
binaries, namely, Cyg X-1 \cite{cui98} and GX~339-4 \cite{bel99}.

The outburst of 1998 was distinctly different from the other two
observed from 4U~1630--47 by RXTE. It had a FRED-type light curve,
with secondary maxima, as typical for X-ray transients 
\cite{chen97}. At the peak the source was in an unusual very high state, 
similar to what has been observed in X-ray Nova XTE~J1748-288 \cite{rtb99},
and in the Galactic microquasars GRS~1915+105 \cite{Morgan97} and 
GRO~J1655-40 \cite{rem99}. So we see that the same 
source 4U~1630--47 demonstrates behavior seen in several Galactic
binaries, which have one in common: all are black hole
candidates.  Observations of 4U~1630-47 confirm that all black
hole systems have some similar X-ray properties, independent
on the type of optical companion (low mass dwarves for X-ray Novae 
or early-type giant for Cyg X-1), and
on long-term flux behavior (persistent, transient or chaoticly
variable).  4U~1630-47 may be considered as an intermediate case
connecting different types of black hole binaries.

\section{CONCLUSIONS}

We analyzed the variability and spectral evolution of X-ray recurrent
transient source 4U~1630--47 during its 1998 outburst. The
RXTE satellite has observed three outbursts from this source to date.  
The behavior of the source during 1998 outburst differed 
significantly from previous outburst in 1996 and from the following 
one in 1999.  While the interval between 1996 and 1998 outbursts was
the same as between other consecutive outbursts observed in 1987--1996
\cite{kuul97}, the 1999 outburst happened much earlier than was expected
\cite{mccol99}.  We suggest that recurrence period of the system 
is not linked with its orbital period, but depends instead 
on the time of matter accumulation in outer accretion disk.

Light curve of 1998 outburst was of FRED-type typical for black
hole X-ray Novae \cite{chen97}. Pointed observations by RXTE provided 
good coverage of all stages of the outburst from the rise of X-ray 
flux until late into the decline.  We concentrated our analysis on 
the rise and maximum phases.

During the rise, the source demonstrated an interesting stage which
can be considered as a transition between low/hard and very high state.
The energy spectrum of the source was hard with an exponential high 
energy cut-off, similar to low state, while its fast variability was 
of the type more typical for VHS. We note that a similar energy 
spectrum was observed during rise phase from X-ray Novae GS/GRS~1124--68 
\cite{ebi94}, KS~1730-312 \cite{trud96}, and for the low luminosity 
state of GRS 1915+105 \cite{Trudolyubov99.1}.

The maximum of the outburst was divided into three plateaus, with
almost constant flux within each plateau, and fast jumps between them.
The spectral and timing parameters were also quite stable for each 
individual plateau, but distinctly different between the plateaus.  
The variability detected on the first plateau is of special interest.
The source exhibits quasi-regular modulations 
with period of $\sim 10 - 20$ s during these observations.
A similar type of variability was observed for both known Galactic 
microquasars GRS~1915+105 \cite{Morgan97} and GRO~J1655-40 
\cite{rem99}. This variability holds for rather narrow range 
of luminosity for each source, which is known fact for the aforementioned 
microquasars, and was confirmed also by our observations for 4U~1630--47. 
Our analysis revealed significant differences in spectral and temporal 
behavior of the source at high and low fluxes during this period of time.
The changes observed can be generally explained 
in the framework of the two--phase model of the accretion flow, involving 
a hot inner comptonization region and surrounding optically thick disk.

The evolution of 4U~1630--47 during the 1998 outburst is analogous 
to what was observed from other X-ray transients, namely, GS/GRS 1124-68 
(Nova Muscae 1991; \cite{ebi94}), KS~1730--312 \cite{bor95,trud96}, 
GRS~1739--278 \cite{bor98}, XTE~J1748-288 \cite{rtb99} 
and XTE~J1550-564 \cite{cui99}.  Near the peak of the outburst 
each showed an unusual very hard state dominated by a hard power law.
This state is an addition to the canonical picture for these systems.

The other two outbursts observed with RXTE, in 1996 and
1999, resemble low-high-low state transitions observed from 
'persistent' Galactic black holes GX~339--4 \cite{bel99} and Cyg X--1
\cite{cui98}. From the long-term flux history one may 
conclude that 4U~1630--47 is an intermediate between transients and 
persistent sources.  As we discussed above, it demonstrates also
states similar to the Galactic microquasars.  Both high-mass binaries,
as Cyg X-1, and low-mass binaries, as X-ray Novae, demonstrate
similar spectral and temporal behavior, with several distinct states
and transitions between them.  We suggest that the 
accretion process and the generation of X-rays 
in all those systems has generic features attributable 
to their black hole nature.

\section{ACKNOWLEDGMENTS}

The research has made use of data obtained through the High Energy
Astrophysics Science Archive Research Center Online Service, provided 
by the NASA/Goddard Space Flight Center.

\label{lastpage}

\end{document}